\documentclass[12pt]{article}
\usepackage{euscript}
\usepackage{amsmath}
\usepackage{amsthm}
\usepackage{amssymb}
\usepackage{latexsym}
\usepackage{amsfonts}
\usepackage{amsbsy}
\usepackage{graphicx,psfrag}
\newcommand{\nc}{\newcommand*}
\newcommand{\rnc}{\renewcommand*}

\nc{\reff}[1]{(\ref{#1})}
\nc{\ts}{\textstyle}
\nc{\ds}{\displaystyle}
\nc{\nn}{\nonumber}
\nc{\bib}{\vspace{-6pt}\bibitem}

\def\half{{\frac{1}{2}}}
\rnc{\Bbb}{\mathbb}
\rnc{\rm}{\text}
\nc{\eps}{\epsilon}
\nc{\vep}{\varepsilon}
\nc{\vp}{\varpi}
\nc{\vpi}{\varphi}
\nc{\vt}{\vartheta}
\nc{\vr}{\varrho}
\nc{\wh}{\widehat}
\nc{\wt}{\widetilde}

\def\cHH{{\mathcal H}}
\def\cHP{{\mathcal H}^P}
\def\cHM{{\mathcal H}^M}
\def\mcH{{\mathcal H}}
\def\cH0M{{\mathcal H}_0^M}

\def\cL{{\mathcal L}}
\def\cN{{\mathcal N}}
\def\BC{{\Bbb C}}

\def\BR{{\Bbb R}}

\def\one{1\hskip-.37em 1}
\def\half{{\frac{1}{2}}}
\def\hW{\wh{W}}

\def\wM{\wt{M}}
\def\hM{\widehat{M}}
\def\wH{\wt{H}}
\def\hH{{\widehat{\mathcal H}}^M}

\def\wX{\wt{X}}
\def\wP{\wt{P}}
\def\CH0M{{\stackrel{\circ\,\,}{\cHM}}}
\def\H0M{{\stackrel{\circ\,\,}{H^M}}}
\def\KPM{\wh{K}_{\pm}^M}
\def\KP{\wh{K}_{+}^M}
\def\KM{\wh{K}_{-}^M}
\def\Kp{\stackrel{\circ\,\,}{K_+^M}}
\def\Km{\stackrel{\circ\,\,}{K_-^M}}
\def\wap{\wt{a}^+}
\def\wam{\wt{a}^-}

\def\hkp{\wh{K}_+}
\def\hkm{\wh{K}_-}

\def\hko{\wh{K}_1}
\def\hkt{\wh{K}_2}
\def\kp{{K}_+^M}
\def\km{{K}_-^M}
\def\kpm{{K}_{\pm}^M}
\def\kz{{K}_0^M}
\def\hkz{\wh{K}_0^M}
\def\ko{{K}_1^M}
\def\kt{{K}_2^M}
\def\wrho{\wt{\rho}}
\nc{\lan}{\langle}
\nc{\ran}{\rangle}
\nc{\ket}[1]{{\vert{#1}\rangle}}                
\nc{\bra}[1]{{\langle{#1}\vert}}                
\nc{\braket}[2]{{\langle{#1}\vert{#2}\rangle}}  
\nc{\vev}[1]{\lan{#1}\ran}                      
\nc{\Szi}[1]{\sum_{ #1 = 0 }^{\infty}}
\nc{\Izi}[2]{\int_{ 0 }^{\infty}{#1}{\rm d}\,{#2}}
\nc{\hgs}[5]{{{}_{#1}{F}_{#2}}%
 \left(\ts{\genfrac{}{}{0pt}{}{#3}{#4}%
 \biggl.\biggr| #5 }\right)}
\nc{\gghf}[3]{{{}_{2}{F}_{1}}%
 \left(\ts{\genfrac{}{}{0pt}{}{#1}{#2}%
 \biggl.\biggr| #3 }\right)}
\nc{\PMP}[1]{P^{\nu}_{#1}(\xi;\,\varphi)}
\nc{\hPMP}[1]{\wh{P}^{\nu}_{#1}(\xi;\,\varphi)}
\def\jpa#1#2#3#4{{\it J. Phys. A.} {\bf #1}, {no.#2}, {#3} {(#4)};}
\def\pra#1#2#3#4{{Phys. Rev. A.} {\bf #1}, {no.#2}, {#3} {(#4);}}
\def\jmp#1#2#3#4{{J. Math. Phys.} {\bf #1}, {no.#2}, {#3} {(#4);}}
\begin{document}

{}
\bigskip

\hspace{10cm}Published in Russian in

\hspace{10cm}{\it Zapski Nauchn. Semin. POMI},
 
\hspace{10cm}{\bf 317}, 66-93 (2004)

\bigskip

\centerline {\large\bf V.V.Borzov, E.V.Damaskinsky}
\bigskip

\centerline {\large\bf The generalized coherent states}
\bigskip
\centerline {\large\bf for oscillators, connected with}
\bigskip
\centerline {\large\bf Meixner and Meixner-Pollaczek polynomials
\footnote {This research is executed at support
  ђ””€ grants No 03-01-00837, 03-01-00593}}

\bigskip
\bigskip
\centerline {\it Authors dedicate this work to our friend and the colleague} 
\bigskip
\centerline {\it P.P.Kulish in connection with his $60^{\rm th}$ birthday}  

\bigskip
\bigskip

\begin{quote}
{\small
The investigation of the generalized coherent states
for oscillator-like systems connected with given family of
orthogonal polynomials is continued. In this work we consider
oscillators connected with Meixner and Meixner-Pollaczek
polynomials and define generalized coherent states for these
oscillators. The completeness condition for these states is
proved by the solution of the related classical moment problem.
The results are compared with the other authors ones. In
particular, we show that the Hamiltonian of the relativistic
model of linear harmonic oscillator can be thought of as the
linearization of the quadratic Hamiltonian which naturally
arised in our formalism.}
\end{quote}

{\small \tableofcontents }

\section{Introduction}

The interest to construction and investigation of coherent
states for oscillator-like systems connected with a family
of orthogonal polynomials has been grown in the late years
(see, for example,~\cite{FlorVin91}-\cite{Appl};
there is the numerous bibliography in
reviews~\cite{AAZM}-\cite{Dodo}).

In works~\cite{bdk}-\cite{Borz00} we suggested a new method for
constructions of oscillator-like systems (or, to put it briefly
"oscillator"), which are connected with a family of orthogonal
polynomials just as the usual boson oscillator with Hermite
polynomials. This approach contains the construction of the
generalized coherent states for such oscillators. In the previous
works we considered the generalized coherent states for
oscillators related to the classical orthogonal polynomials
(Laguerre~\cite{bd4}, Legendre~\cite{bd2}, Chebyshev~\cite{bd3},
Gegenbauer~\cite{bd5}) and the Hermite
q-polynomials~\cite{bd7}-\cite{bd9}.

We describe in brief the essential points of our approach.
Firstly, we put the recurrent relations for these polynomials
in a symmetrical form~\cite{Borz00} (with symmetrical Jacobi
matrix) by a renormalization of these polynomials.
Secondly,
by standard way~\cite{Borz00}
we introduce the (generalized) coordinate and momentum operators,
the associated ladder operators of creation and annihilation and
{\it quadratic} Hamiltonian which spectrum is defined by the
coefficients of (symmetrized) recurrent relations for considered
polynomials. In this way we define the oscillator connected with
given polynomials. Realizing the creation and annihilation
operators for constructed oscillator as a differential (or a
difference) operators (see \cite{bd1}), we get a differential
(or a difference) equation for eigenvectors of the above
quadratic Hamiltonian. As a rule, (for known systems of
orthogonal polynomials) this equation is equivalent to the
standard second order  differential (or difference) equation for
these polynomials. It is natural to consider this equation as an
analogue of the Schr\"odinger equation for oscillator being
investigated.

Next we introduce generalized coherent states of Barut-Girardello
type~\cite{BG-CS}) as the eigenstates of the annihilation operator
for constructed oscillator. To prove the (over)co\-mple\-te\-ness
of this family of states we have solve an appropriate classical
moment problem. The measure which worked out this moment problem
is employing in resolution of unity for the constructed family of
the coherent states. For the classical polynomials (both
depending on a continuous, and on a discrete variable) the arised
classical moment problem is determined and its solution not need
much effort. However for the deformed polynomial systems more
difficult undetermined moment problem sturt up
(see ~\cite{CMP}-\cite{sim97}). In this case we used the results
of research of this problem obtained together with P.P.Kulish
in~\cite{bdk}.

Note that the derived explicit form of coherent states for
generalized oscillators associated with orthogonal polynomials
allows to calculate the values of some physically interesting
quantities (such as, for example, Mandel parameter) for such
systems.

If there is the dynamical symmetry group (or algebra) for
constructed oscillator-like system, it is possible
to define the Perelomov type coherent states in the standard
way~\cite{PerelomovBk} (as action of the unitary shift operator
on the fixed state vector). Recall that for the standard
boson oscillator connected with the Heisenberg group the
Barut--Girardello coherent states  coincide with coherent states
of the Perelomov type, as well as with the states minimizing the
Heisenberg uncertainty relation. However this is not so in the
more complicated cases.

It is essential for possible applications that the suggested
construction of oscillator-like systems and related coherent
states allows to pick up for given energy spectrum a suitable
family of orthogonal polynomials (such that the coefficients of
recurrent relations for these polynomials determine the spectrum
of the Hamiltonian) diagonalising Hamiltonian of this system.
In this way we bypass a difficult factorisation problem for
Hamiltonian of this system.

Further note that together with classical orthogonal polynomials
and their deformed analogues in physical researches of last years
the growing attention is given to discrete polynomials (such as,
for example, Hahn polynomials~\cite{Hahn}-\cite{KS94}),
satisfying not a differential but a difference equation.
After the publication of the works~\cite{AS85}-\cite{Askey},
where the continual analogues of such polynomials (that is the
polynomials which argument is extended to continuous values,
the orthogonality relation is written by an integral, and,
finally, the index is continued in complex plane)
was introduced, the connection of these polynomials
(and also the Meixner and Meixner - Pollaczek
polynomials~\cite{Meixner34}-\cite{Pollaczek}) with the
Heisenberg group was founded~\cite{Bender}-\cite{Koor89}.
So it is natural to investigate the oscillator-like systems
defined by these polynomials. Attempts of such construction '
was undertaken in the work ~\cite{AW-Meixner} (Meixner and
Meixner-Pollaczek polynomials) and in \cite{NJI}
(Hahn polynomials). The Pollaczek polynomials were involved in
the description~\cite{Atak84} of the wave functions of
relativistic model of linear harmonic oscillator in the
frame-work of the quasi-potential approach. (For more details on
this model and its variants we refer the reader
to~\cite{AW89}-\cite{ANW}).

This model was used also in~\cite{AW-Meixner}, where Hamiltonian
(which spectrum depends linearly on $n$) was defined. In this
work for the case of Meixner and Meixner - Pollaczek polynomials
Barut - Girardello  coherent states are constructed and it was
shown that $sp(2,\BR)$ is dynamical symmetry algebra of
considered model and that the Hamiltonian is one of the
generators of this algebra. That allows to define Perelomov type
coherent states for this model.

Because the Meixner and Meixner-Pollaczek  polynomials fulfill
three-terms recurrent relation, the construction of
oscillator-like systems described above is also applied to them.
In the present work we shall construct these oscillator-like
systems and define the generalized coherent states (both
Barut - Girardello type and Perelomov type) for these systems.

 We shall show, in particular, that Hamiltonian of the model,
considered in~\cite{AW-Meixner}, can be thought as a linearization
of quadratic Hamiltonian, naturally arising in our approach.
Note that the connection of the Meixner oscillator (at specific
value of parameter $\varphi=\frac{\pi}{2}$) with the relativistic
model of linear oscillator~\cite{AMN} was mentioned  in the
work~\cite{AW-Meixner}. However this connection was not obvious
and the reasons of appearence of specific value of parameter
$\varphi =\frac{\pi}{2}$ were not clear because of absence of
the appropriate calculations.  In the given work we shall
analyze this connection and show, that only for values of
parameter $\varphi =\frac{\pi}{2}+k\pi$ Hamiltonian of the
Meixner oscillator coincide with Hamiltonian of the
quasi-potential model of relativistic oscillator from~\cite{AS85}.

\section {Meixner oscillator}
\subsection {Meixner polynomials}

Let us remember, that a generalized hypergeometric series
${{}_{r}{F}_{s}}$ defined by
\begin{equation}\label{001}
\hgs{r}{s}{a_1,\ldots,a_r}{b_1,\ldots,b_s}{z}=
\Szi{k}\frac{(a_1)_k (a_2)_k\cdots (a_r)_k}
{(b_1)_k (b_2)_k\cdots (b_s)_k}\,
\frac{z^k}{k!}\,,
\end{equation}
where the shifted factorials (Pochhammer symbols) are given by
\begin{equation}\label{002}
(a)_0=1,\quad
(a)_k=a(a+1)(a+2)\ldots(a+k-1)=\frac{\Gamma(a+k)}{\Gamma(a)}.
\end{equation}

For $\beta >0$ and $0<\gamma<1$ the Meixner
polynomials~\cite{Meixner34}
\begin{equation}\label{003}
M_n(\xi;\beta,\gamma)=
\gghf{-n,-\xi}{\beta}{1-\frac{1}{\gamma}}=
M_{\xi}(n;\beta,\gamma).
\end{equation}
form two-parameter polynomial family
$n=0,1,2, \ldots,$ fulfilling the orthogonality relation
\begin{equation}\label{004}
\Szi{m}\, \rho^{M}(m)\,M_n(m;\beta,\gamma)\,
M_k(m;\beta,\gamma) = d_n\,\delta_{nk},
\end{equation}
with respect to the weight function
\begin{equation}\label{005}
\rho^{M}(m)(\xi)=\frac{(\beta)_\xi\,\gamma^\xi}{\xi!}\,,
\end{equation}
where the value of a square of norm is given by
\begin{equation}\label{006}
d_n=\frac{n!}{\gamma^n(\beta)_n(1-\gamma)^\beta}\,.
\end{equation}
These polynomials also fulfill the recurrent
relations~\cite{Meixner34,KS94}
\begin{equation}\label{007}
\left[n+(n+\beta)\gamma-(1-\gamma)\xi\right]\,
M_n(\xi;\beta,\gamma) =
(n+\beta)\gamma M_{n+1}(\xi;\beta,\gamma)+
nM_{n-1}(\xi;\beta,\gamma)\,.
\end{equation}
The difference equation for Meixner  polynomials looks
like~\cite{Meixner34,KS94}
\begin{equation}\label{008}
[\gamma(\xi+\beta)e^{\partial_\xi}+
\xi e^{-\partial_\xi}-(1+\gamma)(\xi+\half\beta)+
(1-\gamma)(n+\half\beta)]\,M_n(\xi;\beta,\gamma)=0\,.
\end{equation}
The reproducing functions for Meixner polynomials
have the forms
\begin{gather}
\Szi{n}\frac{(\beta)_n}{n!}\,M_{n}(\xi;\beta,\gamma)\,t^n=
\left(1-\frac{t}{\gamma}\right)^{\xi}
\left(1-t\right)^{-\xi-\beta}\,,
\label{009}\\
\Szi{n}\frac{1}{n!}\,M_{n}(\xi;\beta,\gamma)\,t^n=
e^t\,\hgs{1}{1}{-\xi}{\beta}
{\left(\frac{1-\gamma}{\gamma}\right)t}\,.
\label{010}
\end{gather}

To get a symmetrical form of recurrent relations \reff{007}
we (following~\cite{Borz00}) define the renormalized Meixner
polynomials
\begin{equation}\label{012}
\wM_n(\xi;\beta,\gamma)=
\frac{M_n(\xi;\beta,\gamma)}{c_n},
\end{equation}
with
\begin{equation}\label{013}
c_0=1,\qquad c_n=\gamma^{-n/2}\sqrt{\frac{n!}{(\beta)_n}},
\quad n\geq1,
\end{equation}
and where we take into account, that
$M_0(\xi;\beta,\gamma)=1.$
Then polynomials  $\wM_n(\xi;\beta,\gamma)$
satisfy canonical three-term recurrent relations with
symmetrical Jacobi matrix
\begin{gather}
\xi\wM_n(\xi;\beta,\gamma)=b_n\wM_{n+1}(\xi;\beta,\gamma)
-a_n\wM_n(\xi;\beta,\gamma)+b_{n-1}\wM_{n-1}(\xi;\beta,\gamma)
\label{014}\\
\wM_0(\xi;\beta,\gamma)=1\,,
\label{015}
\end{gather}
which coefficients are defined by the formulas
\begin{equation}\label{016}
b_{-1}=0,\quad b_{n}=\frac{\sqrt{\gamma}}{\gamma-1}
\sqrt{(\beta+n)(n+1)},
\quad a_{n}=\frac{n+(n+\beta)\gamma}{\gamma-1},\qquad n\geq0.
\end{equation}
Let us remark that Jacobi matrix defined by the relation
\reff{014} has a nonzero diagonal and the appropriate  moment
problem is determined~\cite{bdk}.

For normalized Meixner polynomials  $\wM_n(\xi;\beta,\gamma)$
the orthogonality relation \reff{004} takes the form
\begin{equation}\label{017}
\Szi{\xi}\,\wrho^{M}(\xi)\,\wM_n(\xi;\beta,\gamma)\,
\wM_m(\xi;\beta,\gamma)=\delta_{nm}
\end{equation}
with the weight function
\begin{equation}\label{018}
\wrho^{M}(\xi)=\frac{(\beta)_\xi\,\gamma^\xi}{\xi!}
(1-\gamma)^{\beta}\,.
\end{equation}
The left hand side of the relation \reff{017}
can be rewritten as integral over the discrete measure with
carriers in the points $\xi=0,1,2,\ldots$ and the loadings
given by \reff{018}.

Now we define Meixner functions
\begin{equation}\label{019}
\psi^{M}_n(\xi;\beta,\gamma)=
(-1)^n\sqrt{\wrho^{M}(\xi)}\,\wM_n(\xi;\beta,\gamma)\,,
\end{equation}
which satisfy in the space $\cH0M:=\ell^2$ the discrete
orthogonality relations
\begin{equation}\label{020}
\Szi{\xi}\,\psi^{M}_n(\xi;\beta,\gamma)\,
\psi^{M}_m(\xi;\beta,\gamma)=\delta_{nm}\,,
\end{equation}
or (in view of symmetry of Meixner polynomials \reff{003})
the orthogonality relations
\begin{equation}\label{022}
\Szi{n}\,\psi^{M}_n(\xi;\beta,\gamma)\,
\psi^{M}_m(\xi';\beta,\gamma)=\delta_{\xi\xi'}\,.
\end{equation}

\subsection{Meixner oscillator}
We will introduce the Hilbert space $\cHM:=\ell^2(\wrho^{M})$
with weight function $\wrho^{M}$ \reff{018}  and the basis
$\left\{\wM_n(\xi;\beta,\gamma)\right\}_{n=0}^{\infty}.$
Let $\hH$ be the same space $\cHM$ with the basis
$\left\{\hM_n(\xi;\beta,\gamma)\right\}_{n=0}^{\infty},$
where
$$\hM_n(\xi;\beta,\gamma)=(-1)^n\wM_n(\xi;\beta,\gamma)\,.$$
The polynomials $\hM_{n}$ satisfy the recurrent relations
\begin{equation}\label{022a}
(-\xi)\hM_n(\xi;\beta,\gamma)=
b_n\hM_{n+1}(\xi;\beta,\gamma)+
a_n\hM_{n}(\xi;\beta,\gamma)+
b_{n-1}\hM_{n-1}(\xi;\beta,\gamma).
\end{equation}

Further we define urther the following unitary transformations
\begin{align}
U_1:\cH0M\rightarrow\hH &\quad\Rightarrow\quad
U_1\varphi=\frac{\varphi}{\sqrt{\wrho^M}}=
\wh{\varphi}\in\hH;\label{023a}\\
U_2:\hH\rightarrow\cHM &\quad\Rightarrow\quad
U_2\hM_n=\wM_n,\quad n=0,1,2,\ldots;\label{023b}\\
U:\cH0M\rightarrow\cHM &\quad\Rightarrow\quad
U=U_2\circ U_1\,.\label{023c}
\end{align}

Following~\cite{Borz00}, we define  "coordinate" $X$ and
momentum $P$ operators by their action on elements of basis
$\left\{\wM_n(\xi;\beta,\gamma)\right\}_{n=0}^{\infty}$
in Hilbert space $\cHM $ according to formulas
\begin{gather}
X\wM_0(\xi;\beta,\gamma)=
b_0\wM_1(\xi;\beta,\gamma)-a_0\wM_0(\xi;\beta,\gamma)\,,\nn\\
X\wM_n(\xi;\beta,\gamma)=
b_n\wM_{n+1}(\xi;\beta,\gamma)-a_n\wM_n(\xi;\beta,\gamma)+
b_{n-1}\wM_{n-1}(\xi;\beta,\gamma),\quad n\geq1;
\label{024}\\
P\wM_0(\xi;\beta,\gamma)=
-ib_0\wM_1(\xi;\beta,\gamma)-a_0\wM_0(\xi;\beta,\gamma)\,,\nn\\
P\wM_n(\xi;\beta,\gamma)=
i\left(b_{n-1}\wM_{n-1}(\xi;\beta,\gamma)
-b_n\wM_{n+1}(\xi;\beta,\gamma)\right)
-a_n\wM_n(\xi;\beta,\gamma),\quad n\geq1. \label{025}
\end{gather}
Now we define oscillator-like system, which we shall name
{\it Meixner oscillator}, by introducing the generalized
coordinate $\wX$
and the generalized momentum $\wP$
\begin{equation}\label{026}
\wX:=\text{Re}(X-P),\qquad \wP:=-i\text{Im}(X-P),
\end{equation}
and creation and annihilation operators
\begin{equation}\label{027}
\wap:=\frac{1}{\sqrt{2}}\left(\wX+i\wP\right),\qquad
\wam:=\frac{1}{\sqrt{2}}\left(\wX-i\wP\right).
\end{equation}
The Hamiltonian of the Meixner oscillator we choose in the
form
\begin{equation}\label{028}
\wH^M=\wX^2+\wP^2\,.
\end{equation}
It is follows from the work~\cite{Borz00} that a spectrum
of the Hamiltonian looks like
\begin{gather}
\lambda_0=2b_0^{\,\,2}=
2\beta\left(\frac{\sqrt{\gamma}}{\gamma-1}\right)^2;
\nn\\
\lambda_n=2\left(b_{n-1}^{\,\,2}+b_{n}^{\,\,2}\right)=
\left(\frac{2\sqrt{\gamma}}{\gamma-1}\right)^2
(n^2+n\beta+\half\beta),\quad n\geq1.\label{029}
\end{gather}
Note that it is possible to define in the Hilbert space
$\wh\cHM$ the Meixner oscillator connected with polynomials
$\wh{M}_n$ by similar considerations. However, from
\reff{024}-\reff{028} it follows that these oscillators
coincide.

In just the same way as in~\cite{Borz00} it is possible to
show that the eigenvalue equation
$\wH^My =\lambda_ny$ is equivalent to the difference equation
\begin{equation}\label{030}
n(\gamma-1)y(\xi)=\gamma(\xi+\beta) y(\xi+1)-
[\xi+(\xi+\beta)\gamma]y(\xi)+\xi y(\xi-1),\qquad
y(\xi)=\wM_n(\xi;\beta,\gamma),
\end{equation}
which it is natural to call the Schr\"odinger equation of
constructed Meixner oscillator. (Note that the equation
\reff{030} is another form of the equation \reff{008}).

We point out one essential difference of our Meixner oscillator
from ones considered in~\cite{AW-Meixner}. The Meixner
oscillator Hamiltonian from~\cite{AW-Meixner} in the space
$\cH0M$ has the following form
\begin{equation}\label{031}
\H0M(\xi)=\frac{1+\gamma}{1-\gamma}\left(\xi+\half\beta\right)-
\frac{\sqrt{\gamma}}{1-\gamma}\left[\mu(\xi)e^{\partial_{\xi}}+
\mu(\xi-1)e^{-\partial_{\xi}}\right],\quad
\mu(\xi)=\sqrt{(\xi+1)(\xi+\beta)}\,.
\end{equation}
Its eigenvalues are linear in $n$
\begin{equation}\label{032}
\mu_n=n+\half\beta,\qquad n=0,1,2,\ldots\,,
\end{equation}
whereas the spectrum of the Hamiltonian  $\wH^M$ has
quadratic dependence on $n$ (see \reff{029}) in our case.
The eigenvalue equation of Hamiltonian $\H0M(\xi)$ \reff{031}
can be written as a difference equation (coinciding with
\reff{008} and \reff{030})
\begin{equation}\label{033}
\left[\gamma(\xi+\beta\,)e^{\partial_{\xi}}+\xi\,
e^{-\partial_{\xi}}
-(1+\gamma)(\xi+\half\beta)+(1-\gamma)(n+\half\beta)\right]
M_n(\xi;\beta,\gamma)=0.
\end{equation}

\section {Dynamic symmetry algebra and connection of
Hamiltonians $\wH$ and $H$}
In the Hilbert space $\cHM$ we shall define operators
\begin{equation}\label{034}
\kp:=\frac{1-\gamma}{\sqrt{2\gamma}}\wap,\quad
\km:=\frac{1-\gamma}{\sqrt{2\gamma}}\wam,\quad
\kz:=\half[ \km,\kp ]\,,
\end{equation}
and define a new Hamiltonian $H^M=U\H0M U^{-1}=\kz.$
It is possible to show that operators $\kp,\km,\kz$
give realization of commutation relations
\begin{equation}\label{035}
[\kz,K_{\pm}^M]=\pm K_{\pm}^M,\qquad [ \km,\kp ]=2\kz;
\end{equation}
of  $sp(2,\BR)$ algebra.

In the work \cite{AW-Meixner} the following realization of
generators $\Kp=U^{-1}\kp U$ and $\Km=U^{-1}\km U$ as
difference operators
(in space $\CH0M$) was given
\begin{gather}
\Kp=\frac{\gamma}{1-\gamma}\mu(\xi)e^{\partial_\xi}+
\frac{1}{1-\gamma}e^{-\partial_\xi}\mu(\xi)-
\frac{2\sqrt{\gamma}}{1-\gamma}\left(\xi+\half\beta\right),
\label{036}\\
\Km=\frac{\gamma}{1-\gamma}\mu(\xi)e^{\partial_\xi}+
\frac{\gamma}{1-\gamma}e^{-\partial_\xi}\mu(\xi)-
\frac{2\sqrt{\gamma}}{1-\gamma}\left(\xi+\half\beta\right),
\label{037}
\end{gather}
Let $\KPM=U_2^{-1}\kpm U_2$ and $\wh{H}^M=U_2^{-1}H^MU_2$ are
the generators of algebra $sp(2,\BR)$ in the space $\hH,$
which are connected to generators from \cite{AW-Meixner}
by unitary transformation
\begin{equation}\label{038}
\KP=U_1{\Kp}U_1^{-1},\quad
\KM=U_1{\Km}U_1^{-1},\quad
\wh{H}^M=U_1{H_0^M}U_1^{-1}.
\end{equation}
From \reff{031}, \reff{036} and \reff{037} it follows that
\begin{gather}
\hkp^M=\frac{\gamma^{3/2}}{1-\gamma}\left(\xi+\beta\right)e^{\partial_\xi}+
\frac{1}{\sqrt{\gamma}(1-\gamma)}\xi e^{-\partial_\xi}-
\frac{2\sqrt{\gamma}}{1-\gamma}\left(\xi+\half\beta\right),
\label{039}\\
\hkm^M=\frac{\sqrt{\gamma}}{1-\gamma}\left(\xi+\beta\right)e^{\partial_\xi}+
\frac{\sqrt{\gamma}}{1-\gamma}\xi e^{-\partial_\xi}-
\frac{2\sqrt{\gamma}}{1-\gamma}\left(\xi+\half\beta\right),
\label{040}\\
\wh{H}^M=\frac{1+\gamma}{1-\gamma}\left(\xi+\half\beta\right)-
\frac{\gamma}{1-\gamma}\left(\xi+\beta\right)e^{\partial_\xi}-
\frac{1}{1-\gamma}\xi e^{-\partial_\xi}\,.
\label{041}
\end{gather}
On elements $\hM_{n}$ of the basis in the space $\hH$ these operators
act by the following formulas
\begin{equation}\label{042a}
\KP\hM_{n}=\mu(n)\hM_{n+1},\quad
\KM\hM_{n}=\mu(n-1)\hM_{n-1},\quad
\wh{H}^M\hM_{n}=\left(n+\half\beta\right)\hM_{n}\,,
\end{equation}
and on elements $\wM_{n}$ of the basis in the space $\wH$ by
\begin{equation}\label{042}
\kp\wM_{n}=-\mu(n)\wM_{n+1},\quad
\km\wM_{n}=-\mu(n-1)\wM_{n-1},\quad
H^M\wM_{n}=\left(n+\half\beta\right)\wM_{n}.
\end{equation}

The cartesian generators $\ko,\,\kt$ of the algebra $sp(2,\BR)$ and the
Cazimir operator ${\mathcal{C}}_2$
are defined by the relations
\begin{gather}
\hko^M=-\frac{i}{2}(\hkp^M - \hkm^M)=\frac{i}{2}\left(
\mu(\xi)e^{\partial_{\xi}}-\mu(\xi)e^{-\partial_{\xi}}\mu(\xi)\right),
\label{043}\\
\hkt^M=-\frac{1}{2}(\hkp^M + \hkm^M)=-\frac{1+\gamma}{2(1-\gamma)}\left(
\mu(\xi)e^{\partial_{\xi}}+\mu(\xi)e^{-\partial_{\xi}}\mu(\xi)\right)
+\frac{2\sqrt{\gamma}}{1-\gamma}(\xi+\half\beta),
\label{044}\\
\wh{\mathcal{C}}_2=(\wh{K}^M)^2=(\hkz)^2-(\hko^M)^2-(\hkt^M)^2=
(\hkz)^2-\hkz-\hkp^M\hkm^M=
\half\beta(\half\beta-I).\label{045}
\end{gather}

Selfadjoint Hamiltonians $\wh{\wH}^M=U_2^{-1}{\wH}^MU_2$ and $H^M$ in the
Hilbert space $\wh{\mathcal{H}}^M$ have the same set of eigenfunctions
$\left\{\wh{M}_n(\xi;\beta,\gamma)\right\}_{n=0}^{\infty}$
and are connected with each other by the relation
\begin{equation}\label{046a}
\wh{\wH}^M=\frac{4\gamma}{(\gamma -1)^2}\left((\wh{H}^M)^2-
(\wh{K}^M)^2\right).
\end{equation}
Hence, the self-adjoint Hamiltonians $\wH^M$ and $H^M$ in the
Hilbert space $\cHM$ are connected with each other by same relation
\begin{equation}\label{046}
\wH^M=\frac{4\gamma}{(\gamma -1)^2}\left((H^M)^2-(K^M)^2\right).
\end{equation}
This formula gives an interesting connection of our Hamiltonian with
Hamiltonian for generalized harmonic oscillator model in the approach used
in~\cite{AW-Meixner}.

\section{Barut - Girardello coherent states for Meixner oscillator}
\subsection {Definition of coherent states}
By definition, Barut - Girardello~\cite{BG-CS} coherent states
for Meixner oscillator are eigenstates of annihilation operator
$\wam.$ Let $\ket{n}=\wM_n(\xi;\beta,\gamma)$ denote the elements of
the Fock basis (the oscillator basis) in Fock Hilbert space (the
space of filling numbers) $\cHH^M.$ Then we have
\begin{equation}\label{047}
\wam\ket{z}=z\ket{z},\qquad \ket{z}=
\cN^{-1}(|z|^2)\Szi{n}\frac{z^n}{\left(\sqrt{2}b_{n-1}\right)!}
\ket{n}\,,
\end{equation}
where coefficients $b_{n}$ are determined by the formula
\reff{016} and we introduce the notation
\begin{equation}\label{048}
\left(\sqrt{2}b_{-1}\right)!=1,\qquad
\left(\sqrt{2}b_{n}\right)!=
2^{\half n}b_{0}b_{1}\cdot\ldots\cdot b_{n-1},\quad n\geq1\,.
\end{equation}

For normalizing factor  $\cN(|z|^2)$ one obtains
\begin{equation}\label{049}
\cN^2(|z|^2)=\braket{z}{z}=
\Szi{n}\frac{|z|^{2n}}{\left(2b_{n-1}^2\right)!}
=\Szi{n}\left(\frac{(1-\gamma)^2}{2\gamma}\right)^n
\frac{|z|^{2n}}{(\beta)_nn!}.
\end{equation}
Because the series in \reff{049} has infinite convergence radius
$R=\infty,$ this series converges on all complex plane.

From the formula \reff{016} it follows that
\begin{equation}\label{050}
\left(2b_{n-1}^2\right)!=
\left(\frac{2\gamma}{(1-\gamma)^2}\right)^n(\beta)_nn!=
\left(\frac{2\gamma}{(1-\gamma)^2}\right)^n
\frac{n!\Gamma(\beta+n)}{\Gamma(\beta)}.
\end{equation}
Taking into account an explicit expression for Bessel function
of the 1-kind
\begin{equation}\label{051}
I_{\alpha}(2\sqrt{z}):
=z^{\half\alpha}\,\Szi{m}\frac{z^m}{m!\Gamma(m+\alpha+1)},
\end{equation}
we receive
\begin{equation}\label{052}
\cN^2(|z|^2)=
\left(\frac{(1-\gamma)^2}{2\gamma}\right)^{\half(\beta-1)}
\frac{\Gamma(\beta)}{|z|^{\beta-1}}\,
I_{\beta-1}(2\frac{(1-\gamma)}{\sqrt{2\gamma}}|z|).
\end{equation}
It allows to rewrite expression \reff{047} for the coherent
state in the following way
\begin{equation}\label{053}
\ket{z}=\frac
{\left(\frac{(1-\gamma)^2}{2\gamma}|z|\right)^{\half(\beta-1)}}
{\left[I_{\beta-1}(2\frac{(1-\gamma)}{\sqrt{2\gamma}}|z|)
\Gamma(\beta)\right]^{\half}}
\Szi{n}\frac{M_n(\xi;\beta,\gamma)}{n!}
\left(\frac{1-\gamma}{\sqrt{2\gamma}}\,z\right)^n.
\end{equation}
Using the reproducing function \reff{010} for Meixner polynomials,
we can evaluate the series in the formula \reff{053}. This allows
to obtain explicit expression for Barut - Girardello coherent
state of Meixner oscillator
\begin{equation}\label{054}
\ket{z}=\sqrt{\frac{1-\gamma}{\sqrt{2\gamma}}}
|z|^{\half(\beta-1)}
\left[\Gamma(\beta)\,
I_{\beta-1}(2\frac{1-\gamma}{\sqrt{2\gamma}}|z|)
\right]^{-\half}\,\exp\left[{\ds\frac{1-\gamma}
{\sqrt{2\gamma}}\,z}\right]\,
\hgs{1}{1}{-\xi}{\beta}
{\left(\frac{(1-\gamma)^2}{\gamma\sqrt{2\gamma}}\right)z}\,.
\end{equation}
For overlapping of two coherent states we obtain
\begin{equation}\label{055}
\braket{z_1}{z_2}=
{I_{\beta-1}(2\frac{1-\gamma}{\sqrt{2\gamma}}
\,\sqrt{\overline{z_1}z_2})}
\left[I_{\beta-1}(2\frac{1-\gamma}{\sqrt{2\gamma}}|z_1|)\,
I_{\beta-1}(2\frac{1-\gamma}{\sqrt{2\gamma}}
|z_2|)\right]^{-\half}\,.
\end{equation}
Note, that for $\alpha=\beta-1$ and $\gamma=2-\sqrt{3}$
this expression coincides with similar one for coherent states
of Laguerre oscillator (see the formula (24) from \cite{bd6}).
Note also that taking into account the relation \reff{018} for
Meixner function, we obtain
\begin{equation}\label{056}
\psi^{M}_n(\xi;\beta,\gamma)=(-1)^n\psi^{}_0(\xi;\beta,\gamma)\,
\wM_n(\xi;\beta,\gamma), \quad
\psi^{M}_0(\xi;\beta,\gamma)=
\sqrt{\frac{(\beta)_{\xi}
{\gamma}^{\xi}}{{\xi}!}(1-\gamma)^{\beta}}\,.
\end{equation}
So our expression for the coherent state is in agreement
with the relation (69) from Atakishiev a.o.
work~\cite{AW-Meixner}, (if we take into account, that in
work~\cite{AW-Meixner} are considered not normalized coherent
states).

\subsection{Proof of the (over)completeness for constructed
family of coherent states}
The most important property of the family of coherent states
is the (over)completeness  property that can be expressed as
validity of the resolution of unity relation
\begin{equation}\label{057}
\iint_{\BC}\ket{z}\bra{z}\, \hW(|z|^2)\text{d}^2z=\one\,.
\end{equation}
To check this formula it is necessary to construct a measure
\begin{equation}\label{058}
\text{d}\mu(|z|^2)=\hW(|z|^2)\text{d}^2z.
\end{equation}
It is known~\cite{GK}~-~\cite{AGMKP} that for this purpose we
have to solve the Stieltjes classical moment
problem~\cite{CMP, sim97}
\begin{equation}\label{059}
\pi\Izi{x^nW(x)}{x}=
\left(\frac{2\gamma}{(1-\gamma)^2}\right)^n\,
\frac{n!\Gamma(\beta+n)}{\Gamma(\beta)},\quad n\geq0,
\end{equation}
where
\begin{equation}\label{060}
W(x)=\frac{\hW(x)}{\cN^2(x)},\quad (x=|z|^2)\,.
\end{equation}
Taking into account the formula (6.561.16) from \cite{G-R},
we find
\begin{equation}\label{061}
W(x)=\frac{2}{\pi}\,\frac{(1-\gamma)^2}{2\gamma\Gamma(\beta)}\,
\left(\frac{(1-\gamma)^2}{2\gamma}\,x\right)^{\half(\beta-1)}
K_{\beta-1}(2\frac{1-\gamma}{\sqrt{2\gamma}}\sqrt{x})\,.
\end{equation}
Then
\begin{equation}\label{062}
\hW(|z|^2)=\frac{(1-\gamma)^2}{\pi\gamma}\,
K_{\beta-1}(2\frac{1-\gamma}{\sqrt{2\gamma}}|z|)\,
I_{\beta-1}(2\frac{1-\gamma}{\sqrt{2\gamma}}|z|)\,,
\end{equation}
and we obtain for a measure \reff{058} the expression
\begin{equation}\label{063}
\text{d}\mu(|z|^2)=\frac{(1-\gamma)^2}{\pi\gamma}\,
K_{\beta-1}(2\frac{1-\gamma}{\sqrt{2\gamma}}|z|)\,
I_{\beta-1}(2\frac{1-\gamma}{\sqrt{2\gamma}}|z|)\,\text{d}^2z.
\end{equation}
For $\alpha=\beta-1$ and $\gamma=2-\sqrt{3}$ this result
coincides with similar expression for the case of Laguerre
polynomials~\cite{bd6}.

\section{The Perelomov coherent states for Meixner oscillator}
The Perelomov coherent states connected with dynamical algebra
$su(1|1)$ in a case of the  Meixner oscillator can be defined by
\begin{equation}\label{071}
\ket{\zeta}=\left(1-|\zeta|^2\right)^{\half\beta}\,
\exp(\zeta\kp)\wM_0(\xi,\beta,\gamma)=
\left(1-|\zeta|^2\right)^{\half\beta}\,
\Szi{n}\sqrt{\frac{(\beta)_n}{n!}}\wM_n(\xi,\beta,\gamma)\zeta^n,
\end{equation}
where $\zeta\in\BC$ and $|\zeta|<1.$\,
Using the reproducing function \reff{010} for Meixner polynomials
and taking into account relations \reff{012}-\reff{013}, we find
\begin{align}
\ket{\zeta}&=\left(1-|\zeta|^2\right)^{\half\beta}\,
\Szi{n}\frac{(\beta)_n}{n!}M_n(\xi,\beta,\gamma)
(\sqrt{\gamma}\zeta)^n \nonumber\\
{}&=\left(1-|\zeta|^2\right)^{\half\beta}\,
\left(1-\frac{\sqrt{\gamma}\zeta}{\gamma}\right)^{\xi}\,
\left(1-\sqrt{\gamma}\zeta\right)^{\xi-\beta}\,.\label{072}
\end{align}
This result coincides with the relation (75)
in \cite{AW-Meixner}. For overlapping of coherent states
\reff{072} we receive the expression
\begin{equation}\label{073}
\braket{\zeta_1}{\zeta_2}=
\left[(1-|\zeta_1|^2)(1-|\zeta_2|^2)\right]^{\half\beta}\,
\left(1-\overline{\zeta_1}\zeta_2\right)^{-\beta}\,,
\end{equation}
which also coincides with  the similar result
from~\cite{AW-Meixner}.

To prove the overcompleteness property it is necessary to
find a measure from resolution of unity for these states. It
brings us to a moment problem
\begin{equation}\label{074}
\pi\Izi{x^nW(x)}{x}=\frac{n!}{(\beta)_n}\,,
\qquad n=0,1,\ldots\,.
\end{equation}
Using the formula (3.251.1) from \cite{G-R}, we receive
\begin{equation}\label{075}
W(x)=\frac{\beta-1}{\pi}(1-x)^{\beta-2}.
\end{equation}
Hence the measure \reff{058} looks like
\begin{equation}\label{076}
\text{d}\mu(|\zeta|^2)=
\frac{\beta-1}{\pi}\frac{{\rm d}^2\zeta}{(1-|\zeta|^2)^2}.
\end{equation}
For $\beta=\alpha+1$ this result also conform with the
similar relation for the case of Laguerre polynomials.

We stress that argument $\zeta$ of Perelomov coherent states
belongs to an interior of unit circle on complex plane
$|\zeta|<1,$ where as in the case of Barut - Girardello
coherent states one has $z\in\BC.$

\section{Meixner - Pollaczek oscillator and its coherent states}
\subsection{Meixner - Pollaczek polynomials}
Meixner - Pollaczek polynomials \cite{KS94} are defined by
hypergeometric function:
\begin{equation}\label{077}
\PMP{n}=\frac{(2\nu)_n}{n!}\,e^{in\varphi}
\gghf{-n,\nu+i\xi}{2\nu}{1-e^{-2i\varphi}},
\end{equation}
and connected with Meixner  polynomials~\reff{003} by
the relation
\begin{equation}\label{078}
\PMP{n}=\frac{e^{-in\varphi}}{n!}(2\nu)_n
M_n(i\xi-\nu;2\nu,e^{-2i\varphi}).
\end{equation}
The orthogonality relation for Meixner - Pollaczek polynomials
has the form  $(\nu>0,\, 0<\varphi<\pi)$
\begin{equation}\label{079}
\frac{1}{2\pi}\int^{\infty}_{-\infty}
e^{(2\varphi-\pi)\xi}|\Gamma(\nu+i\xi)|^2\PMP{m}
\PMP{n}\text{d}\varphi=
\frac{\Gamma(n+2\nu)}{(2\sin\varphi)^{2\nu}n!}\delta_{m,n}.
\end{equation}
These polynomials satisfy the recurrent relations
\begin{equation}\label{081}
\xi \PMP{n}=\wt{b}_n\PMP{n+1}-\wt{a}_n\PMP{n}+\wt{c}_n
\PMP{n-1}; \quad \PMP{0}=1,
\end{equation}
where
\begin{equation}\label{082}
\wt{a}_n=(n+\nu)\text{ctg}\varphi,\quad
\wt{b}_{n}=\frac{n+1}{2\sin\varphi},\quad
\wt{c}_n=\frac{n+2\nu-1}{2\sin\varphi},
\end{equation}
and have the following symmetry property
\begin{equation}\label{080}
P^{\nu}_{n}(-\xi;\,-\varphi)=\PMP{n},
\end{equation}

Renormalized (according~\cite{Borz00}) polynomials
\begin{equation}\label{083}
\hPMP{n}=\PMP{n}\sqrt{\frac{n!}{(2\nu)_n}},\quad n\geq0;
\end{equation}
satisfy the symmetric recurrent relations
\begin{equation}\label{084}
\xi \hPMP{n}=\alpha_n\hPMP{n+1}-\wt{a}_n\hPMP{n}+
\alpha_{n-1}\hPMP{n-1};
\quad \hPMP{0}=1,
\end{equation}
where
\begin{equation}\label{085}
\alpha_n=\frac{\sqrt{(n+1)(2\nu+n)}}{2\sin\varphi}.
\end{equation}

For renormalized Meixner - Pollaczek polynomials the
orthogonality relation becomes
\begin{equation}\label{086}
\int^{\infty}_{-\infty}\hPMP{m}\hPMP{n}\mu(\text{d}\xi)
=\delta_{m,n},
\end{equation}
where
\begin{equation}\label{087}
\mu(\text{d}\xi)=\frac{|\Gamma(\nu+i\xi)|^2}
{2\pi\Gamma(2\nu)}
e^{(2\varphi-\pi)\xi}(2\sin\varphi)^{2\nu}\text{d}\xi
\end{equation}
Let $\cHP$ be the Hilbert space
\begin{equation}\label{088}
\cHP=L^2(\BR,\mu(\text{d}\xi)).
\end{equation}
In what follows we shall consider unitary transformation
$\wh{V}:{\wh{\mcH}}^M\rightarrow\cHP.$ Note that
the parameter $\gamma$ included in recurrent relations
\reff{014} and \reff{022a} (for $\wt{M}_n$ and $\wh{M}_n,$
accordingly) is equal to
$e^{-2i\varphi}.$ So for $\sqrt{\gamma}=\sqrt{e^{-2i\varphi}}$
it is possible to choose two values
\begin{equation}\label{089}
\pm\sqrt{\gamma}=\mp e^{-i\varphi}.
\end{equation}
From \reff{022a} choosing a minus sign, we obtain a recurrent
relations
\begin{equation}\label{090}
\xi\hM_n^-(\xi;\beta,\gamma)=
-i\alpha_n\hM_{n+1}^-(\xi;\beta,\gamma)-
(i\wt{a}_n+\nu)\hM_{n}^-(\xi;\beta,\gamma)-
i\alpha_{n-1}\hM_{n-1}^-(\xi;\beta,\gamma);
\end{equation}
and from \reff{014} choosing a plus sign, we obtain a recurrent
relations
\begin{equation}\label{091}
\xi\hM_n^+(\xi;\beta,\gamma)=
i\alpha_n\hM_{n+1}^+(\xi;\beta,\gamma)-
(i\wt{a}_n+\nu)\hM_{n}^+(\xi;\beta,\gamma)+
i\alpha_{n-1}\hM_{n-1}^+(\xi;\beta,\gamma).
\end{equation}
These relations differs by a choice of value
$\sqrt{e^{-2i\varphi}},\quad(0<\varphi<\pi).$

Taking into account \reff{012}, from \reff{078} it follows that
$$
\wh{P}_n^{\nu}(\xi,\varphi)=\wt{M}_n^-
(i\xi-\nu,2\nu,e^{-2i\varphi}).
$$
Than we can define unitary transformation $\wh{V}$ by
the relation
\begin{equation}\label{092}
\wh{P}_n^{\nu}(\xi,\varphi)=
\wt{M}_n^-(i\xi-\nu,2\nu,e^{-2i\varphi})=
(-1)^n\wh{M}_n^-(i\xi-\nu,2\nu,e^{-2i\varphi})=
\wh{V}\wh{M}_n^-(\xi,2\nu,e^{-2i\varphi})
\end{equation}
Note that we choose in \reff{089} a minus sign, appropriate
to polynomials $\wh{M}_n^-,$ which satisfy recurrent relations
\reff {090}, because the choice of a plus sign (i.e.
polynomials $\wh{M}_n^+,$ satisfying the recurrent
relations \reff{091}) implies a contradiction. Indeed, in
this case from \reff{091} follows
$\wh{M}_n^+(\xi,2\nu,e^{-2i\varphi})=
\wt{M}_n^-(\xi,2\nu,e^{-2i\varphi})$
and then
$$
\wh{V}\wh{M}_n^+(\xi,2\nu,e^{-2i\varphi})=
(-1)^n\wt{M}_n^-((i\xi-\nu),2\nu,e^{-2i\varphi})\neq
\wt{M}_n^-((i\xi-\nu),2\nu,e^{-2i\varphi}).
$$

It is relevant to note that transition from
discrete orthogonality relations \reff{017}
for Meixner polynomials to continuous orthogonality relations
\reff{086} for Meixner - Pollaczek polynomials is analogues
to Zommerfeld - Watson transformation in the scattering theory.
To proving the unitarity of transformation $\wh{V}$ we
substitute \reff{092} in orthogonality relation \reff{086}
\begin{equation}\label{095}
I=\int^{\infty}_{-\infty}\wh{M}_{n}
(\pm i\xi -\nu;2\nu;e^{-2i\varphi})
\wh{M}_{m}(\pm i\xi -\nu;2\nu;e^{-2i\varphi})
\mu(\text{d}\xi)=\delta_{m,n}.
\end{equation}

\begin{center}
\unitlength=1mm
\special{em:linewidth 0.4pt}
\linethickness{0.4pt}
\begin{picture}(70.00,86.00)
\put(40.00,10.00){\vector(0,1){75.00}}
\put(10.00,75.00){\vector(1,0){60.00}}
\put(40.00,63.00){\circle{1.00}}
\put(40.00,50.00){\circle{1.00}}
\put(40.00,40.00){\circle{1.00}}
\put(40.00,42.00){\oval(10.00,54.00)[t]}
\put(35.00,42.00){\line(0,-1){30.00}}
\put(35.00,20.00){\vector(0,1){8.00}}
\put(45.00,12.00){\line(0,1){30.00}}
\put(45.00,30.00){\vector(0,-1){8.00}}
\put(44.00,86.00){\makebox(0,0)[cc]{$\Im{z}$}}
\put(70.00,78.00){\makebox(0,0)[cc]{$\Re{z}$}}
\put(40.00,30.00){\circle{1.00}}
\put(44.00,63.00){\makebox(0,0)[cc]{\scriptsize $-i\nu$}}
\put(48.00,50.00){\makebox(0,0)[cc]{\scriptsize $-i(\nu+1)$}}
\put(48.00,40.00){\makebox(0,0)[cc]{\scriptsize $-i(\nu+2)$}}
\put(48.00,30.00){\makebox(0,0)[cc]{\scriptsize $-i(\nu+3)$}}
\put(20.00,76.00){\vector(1,0){10.00}}
\put(26.00,78.00){\makebox(0,0)[cc]{$C$}}
\put(32.00,48.00){\makebox(0,0)[cc]{$C_2$}}
\end{picture}
\end{center}

Note that $|\Gamma(\nu+i\xi)|^2=
\Gamma(\nu+i\xi)\Gamma(\nu-i\xi)$
and integrand in the left part of the formula \reff{095}
is analytical function  on all complex plane except the points
$\zeta_k=-i(k+\nu),$ \quad $(k\geq0)$
on the imaginary axis in which it has poles of the first order.
Taking into account the asymptotic behaviour of Gamma-function
at infinity we can replace a contour of integration $C_2$ by a
contour $\cL=\bigcup\limits_{k=0}^{\infty}l_k,$
where $l_k$ - a circle with the centre in a point $\zeta_k$
and a radius $r\leq\half.$ Then, applying the residue theorem,
we received
\begin{equation}\label{096}
I=\sum_{k=0}^{\infty}\,\oint_{l_k}F(\zeta;\nu;\varphi)
\text{d}\varphi
=2\pi i\sum_{k=0}^{\infty}\text{Res}F\,
\rule[-8pt]{.2mm}{18pt}_{\,\zeta=\zeta_k}\,.
\end{equation}
Because
\begin{equation}\label{097}
\text{Res}\left(\Gamma(\nu-i\zeta)\right)\,
\rule[-8pt]{.2mm}{18pt}_{\,\zeta=\zeta_k}=
\frac{(-1)^k}{k!}i,\qquad k=0,1,\ldots \,,
\end{equation}
and
other factors in $F$ are analytical
functions,  we obtain \reff{017} from \reff{096}.
Thus unitarity of transformation $V^{-1}:
\cHH_P\rightarrow\cHH_M$
is proved (compare with \cite{AS85}).

\subsection{Meixner - Pollaczek oscillator}

As above, we define in the Hilbert space
$\cHP:=\text{L}^2(\BR,\mu(\text{d}\xi)),$  considered as
a Fock space, coordinate and momentum operators, ladder
operators $\wt{a}^-$ and $\wt{a}^+,$ and also Hamiltonian
$\wt {H}^P.$ Spectrum of the Hamiltonian $\wt{H}^P$ is given
by
\begin{equation}\label{098}
\lambda_0=\frac{\nu}{2\sin^2\varphi},\qquad
\lambda_n=\frac{n(n+2\nu)}{\sin^2\varphi}.
\end{equation}
Note that the eigenvalues equation $\wt{H}^Py=\lambda y$
for the operator $\wt{H}^P$ is equivalent to the difference
equation for Meixner - Pollaczek polynomials
\begin{equation}\label{099}
e^{i\varphi}(\nu-i\xi)y(\xi+i)+
2i(\xi\cos\varphi-(n+\nu)\sin\varphi)y(\xi)-
e^{-i\varphi}(\nu+i\xi)y(\xi-i)=0.
\end{equation}

As above, we can introduce another Hamiltonian, connected
with relativistic oscillator \cite{AS85}. For this purpose
in the Hilbert space $\cHH^P$ we define operators
\begin{equation}\label{100}
K_+^P=\sqrt{2}\sin\varphi\wt{a}^+_P,\quad
K_-^P=\sqrt{2}\sin\varphi\wt{a}^-_P,\quad
K_0^P=\half [K_-^P, K_+^P]=H^P,
\end{equation}
which fulfill the commutation relations of $sp(2,\BR)$ algebra
\begin{equation}\label{101}
[K_-^P, K_+^P]=2K_0^P,\quad [K_0^P, K_{\pm}^P]=\pm K_{\pm}^P.
\end{equation}
The Casimir operator takes the form
\begin{equation}\label{102}
K^2=(K_0^P)^2-(K_1^P)^2-(K_2^P)^2,
\end{equation}
where
\begin{equation}\label{103}
K_1^P=-\frac{i}{2}(K_+^P-K_-^P),\quad K_2^P=
-\frac{i}{2}(K_+^P+K_-^P).
\end{equation}

Using the results of work \cite{Borz00}, we receive from
relations \reff{100}, \reff{034}, \reff{042} and \reff{078}

\begin{equation}\label{104}
K_+^{P}=\wh{V}\wh{K}_+^M\wh{V}^{-1},\quad
K_-^{P}=\wh{V}\wh{K}_-^M\wh{V}^{-1},\quad
H^{P}=\wh{V}\wh{H}^M\wh{V}^{-1}.
\end{equation}
Recall that from \reff{085} and \reff{016} for
$\gamma=e^{-2i\varphi}$ Ё $\beta=2\nu$ we obtain
\begin{equation}\label{104a}
b_n^{\pm}=\mp i\alpha_n,\qquad a_n^{\pm}=i\wt{a}_n+\nu\,.
\end{equation}
Then from recurrent relations \reff{014} and \reff{090} we
have
\begin{gather}
\left( \wh{V}\xi\wh{V}^{-1} \right) \hPMP{n}
   =\wh{V}\xi\wh{M}_n^-(\xi,2\nu,e^{-2i\varphi})= \nn\\
=\wh{V}\left(
-i\alpha_n\wh{M}_{n+1}^-(\xi,2\nu,e^{-2i\varphi})
-(i\wt{a}_n+\nu)\wh{M}_{n}^-(\xi,2\nu,e^{-2i\varphi})
-i\alpha_{n-1}\wt{M}_{n-1}^-(\xi,2\nu,e^{-2i\varphi})
\right)= \nn\\
=-i\left(
\alpha_n\hPMP{n+1}-\wt{a}_n\hPMP{n}+\alpha_{n-1}\hPMP{n-1}
\right)-\nu\hPMP{n}\,.
\label{104c}
\end{gather}
From \reff{104c} and \reff{081} it follows that (for any values
of $\varphi$) the operator which is unitary equivalent to the
"coordinate" operator in the space $\mcH^P$  acts in not natural
way (it is not the operator of multiplication on independent
variable). Only for
$\xi=\frac{\pi}{2}+k\pi$ we have
\begin{equation}\label{104b}
\wh{V}\xi\wh{V}^{-1}=(-i\xi-\nu).
\end{equation}
Therefore in what follows we shall consider the operator
$\wh{V}$ only with $\xi=\frac{\pi}{2}.$ Further we have
\begin{gather*}
\wh{V}e^{\partial_\xi}\wh{V}^{-1}\wh{P}_n^{\nu}(\xi,\varphi)
=\wh{V}e^{\partial_\xi}\wh{M}_n^-(\xi,2\nu,e^{-2i\varphi})\\
=\wh{V}\wt{M}_n^-(\xi+1,2\nu,e^{-2i\varphi})
=(-1)^n\wt{M}_n^-(i(\xi+1)-\nu,2\nu,e^{-2i\varphi})\\
=(-1)^ne^{i\partial_\xi}\wt{M}_n^-(i\xi-\nu,2\nu,e^{-2i\varphi})
=e^{i\partial_\xi}\wh{P}_n^{\nu}(\xi,\varphi).
\end{gather*}
Similarly, we receive
$$
\wh{V}e^{-\partial_\xi}\wh{V}^{-1}=e^{-i\partial_\xi}.
$$
Finally, we have
\begin{equation}\label{105}
\wh{V}\xi\wh{V}^{-1}=-i\xi-\nu,\qquad
\wh{V}e^{\pm\partial_\xi}\wh{V}^{-1}=e^{\pm i\partial_\xi}.
\end{equation}
From these relations, taking into account
\reff{039} - \reff{042}, we get an explicit realization of
operators $K_+^{P},$ $K_-^{P},$ $H^{P}$ as difference operators
in $\cHP$
\begin{gather}
k_+^{P}=K_+^{P}\rule[-12pt]{0.5pt}{22pt}_{\,\varphi=
\frac{\pi}{2}}=\frac{i}{2}(\nu-i\xi)e^{i\partial_\xi}-
\frac{i}{2}(\nu+i\xi)e^{-i\partial_\xi}+\xi,
\label{112}\\
k_-^{P}=K_-^{P}\rule[-12pt]{0.5pt}{22pt}_{\,\varphi=
\frac{\pi}{2}}=-\frac{i}{2}(\nu-i\xi)e^{i\partial_\xi}+
\frac{i}{2}(\nu+i\xi)e^{-i\partial_\xi}+\xi,
\label{113}\\
h^{P}=H^{P}\rule[-12pt]{0.5pt}{22pt}_{\,\varphi=\frac{\pi}{2}}=
\frac{\nu-i\xi}{2}e^{i\partial_\xi}+
\frac{\nu+i\xi}{2}e^{-i\partial_\xi}\,.
\label{114}
\end{gather}

These operators act on normalized Meixner - Pollaczek
polynomials according to
\begin{equation}\label{115}
K_+^{P}\wh{P}^{\nu}_{n}=-\mu(n)\wh{P}^{\nu}_{n+1},\qquad
K_-^{P}\wh{P}^{\nu}_{n}=-\mu(n-1)\wh{P}^{\nu}_{n-1},\qquad
H^{P}\wh{P}^{\nu}_{n}=(n+\nu)\wh{P}^{\nu}_{n}.
\end{equation}

 To compare our results with the ones from the work of
N.M.Atakishiev and S.K.Suslov \cite{AS85}, we introduce three
auxiliary spaces.
First of them - the Hilbert space
${\stackrel{\circ\,}{\cHH^P}}=\tt{L}^2(\text{d}\xi)$ with
the basis
\begin{equation}\label{116}
\left\{\Phi_n^{\nu}(\xi,\varphi)
=g(\nu;\xi)\wh{P}^{\nu}_{n}(\xi,\varphi)\right\}_{n=0}^{\infty},
\end{equation}
where
\begin{equation} \label{117}
g(\nu;\xi)=\frac{|\Gamma(\nu-i\xi)|2^{\nu}}{\sqrt{2\pi\Gamma(n)}}.
\end{equation}
Using the unitary operator
\begin{equation}\label{118}
W:\cHP\rightarrow{\stackrel{\circ\,}{\cHH^P}},\qquad
W{P}^{\nu}_{n}=g\wh{P}^{\nu}_{n}=\Phi_n^{\nu},
\end{equation}
and inverse of it
\begin{equation}\label{119}
W^{-1}:{\stackrel{\circ\,}{\cHH^P}}\rightarrow\cHP,\qquad
W^{-1}\Phi_n^{\nu}=g^{-1}\Phi_n^{\nu}={P}^{\nu}_{n}\,,
\end{equation}
we define operators
\begin{equation}\label{120}
{\stackrel{\circ\,}{K_+^{P}}}=WK_+^{P}W^{-1},\qquad
{\stackrel{\circ\,}{K_-^{P}}}=WK_-^{P}W^{-1},\qquad
{\stackrel{\circ\,}{H^{P}}}=WH^{P}W^{-1}.
\end{equation}
Because
\begin{equation}\label{121}
We^{i\partial_\xi}W^{-1}=|{\nu-1+i\xi}|e^{i\partial_\xi},\qquad
We^{-i\partial_\xi}W^{-1}=\frac{1}{|\nu+i\xi|}e^{-i\partial_\xi},
\end{equation}
and taking into account  relations \reff{112}-\reff{114},
we obtain the explicit form for operators \reff{120} at
$\varphi=\frac{\pi}{2}$
\begin{gather}
k_+^{P}={\stackrel{\circ\quad}{K_+^{P}}}
\rule[-12pt]{0.5pt}{22pt}_{\,\varphi=\frac{\pi}{2}}=
\frac{i}{2}(\nu-i\xi)|\nu-1+i\xi|e^{i\partial_\xi}-
\frac{i}{2}\frac{\nu+i\xi}{|\nu+i\xi|}e^{-i\partial_\xi}+\xi,
\label{128}\\
k_-^{P}={\stackrel{\circ\quad}{K_-^{P}}}
\rule[-12pt]{0.5pt}{22pt}_{\,\varphi=\frac{\pi}{2}}=
-\frac{i}{2}(\nu-i\xi)|\nu-1+i\xi|e^{i\partial_\xi}+
\frac{i}{2}\frac{\nu+i\xi}{|\nu+i\xi|}e^{-i\partial_\xi}+\xi,
\label{129}\\
h^{P}={\stackrel{\circ\quad}{H^{P}}}
\rule[-12pt]{0.5pt}{22pt}_{\,\varphi=\frac{\pi}{2}}=
\frac{1}{2}(\nu-i\xi)|\nu-1+i\xi|e^{i\partial_\xi}+
\frac{1}{2}\frac{\nu+i\xi}{|\nu+i\xi|}e^{-i\partial_\xi}.
\label{130}
\end{gather}

Let us denote by ${\stackrel{\circ\,}{\cHH_A^P}}$ the space
${\stackrel{\circ\,}{\cHH^P}}$
with the following choice of parameters
\begin{equation}\label{131}
\varphi=\frac{\pi}{2},\qquad \xi=\frac{x}{\lambda},\qquad
\nu(\nu-1)={\lambda}^{-4}\,.
\end{equation}
Let us choose in the space ${\stackrel{\circ\,}{\cHH_A^P}}$
a new basis
\begin{equation}\label{132}
{\stackrel{\circ\,}{\Psi_n}}=S\Phi^{\nu}_n=
\Phi^{\nu}_n\,e^{i\text{arg}\Gamma(\nu+i\xi)},
\quad n=0,1,2,\ldots.
\end{equation}
By the action of the unitary operator
$S=e^{i\text{arg}\Gamma(\nu+i\xi)}$ the
difference operators $e^{\pm i\partial_\xi}$  are transformed
according to
\begin{gather}
Se^{i\partial_\xi}S^{-1}=
e^{i\text{arg}\Gamma(\nu+i\xi)}e^{i\partial_\xi}
e^{-i\text{arg}\Gamma(\nu+i\xi)}=
\qquad\qquad\qquad\qquad\qquad \nonumber\\
\qquad\qquad\qquad\qquad=
e^{i\text{arg}(\nu-1+i\xi)}e^{i\text{arg}\Gamma(\nu+i\xi)}
e^{-i\text{arg}\Gamma(\nu+i\xi)}e^{i\partial_\xi}=
e^{i\text{arg}(\nu-1+i\xi)}e^{i\partial_\xi};\label{133}\\
Se^{-i\partial_\xi}S^{-1}=
e^{i\text{arg}\Gamma(\nu+i\xi)}e^{-i\partial_\xi}
e^{-i\text{arg}\Gamma(\nu+i\xi)}=
e^{i\text{arg}(\nu-i\xi)}e^{-i\partial_\xi}.\label{134}
\end{gather}

Then Hamiltonian ${\stackrel{\circ\quad}{H^{P}}}$
\reff{130} became
\begin{equation}
{\stackrel{\circ\quad}{h_S^{P}}}=
S{\stackrel{\circ\quad}{h^{P}}}S^{-1}=
\half\left(\nu(\nu-1)+i\xi+\xi^2\right)e^{i\partial_\xi}+
\half e^{-i\partial_\xi}. \label{137}
\end{equation}
In  the space ${\stackrel{\circ\,}{\cHH_A^P}}$
Hamiltonian \reff{137} takes the form
\begin{equation}\label{138}
{\stackrel{\circ\quad}{h_A^{P}}}=\half
\left(\frac{1}{\lambda^4}+i\frac{x}{\lambda}+
\frac{x^2}{\lambda^2}\right)
e^{i\lambda\partial_x}+\half e^{-i\lambda\partial_x}.
\end{equation}

To check that our Hamiltonian coincides with Hamiltonian from
the work \cite{AS85} it is necessary to pass to the space
${\cHH_A^P}$ by unitary transformation
$V_A:{\stackrel{\circ\,}{\cHH_A^P}}\rightarrow{\cHH_A^P},$
such that
\begin{equation}\label{139}
\phi^A_n=V_A\stackrel{\circ\,}{\phi_n}=
\eta(x)\stackrel{\circ\,}{\phi_n},
\end{equation}
where
\begin{equation}\label{140}
\eta(x)=\lambda^{2i\frac{x}{\lambda}-\half},\quad
\stackrel{\circ\,}{\phi_n}=
\Phi_n^{\nu}(\frac{x}{\lambda},\frac{\pi}{2})
e^{i\text{arg}\Gamma(\nu+i\frac{x}{\lambda})}.
\end{equation}
Under this transformation we have
\begin{gather}
V_Ae^{i\lambda\partial_x}V_A^{\,-1}=
\eta(x)e^{i\lambda\partial_x}\eta^{-1}(x)=
\frac{\eta(x)}{\eta(x+i\lambda)}e^{i\lambda\partial_x}=
\lambda^2e^{i\lambda\partial_x};
\label{141}\\
V_Ae^{-i\lambda\partial_x}V_A^{\,-1}=
\eta(x)e^{-i\lambda\partial_x}\eta^{-1}(x)=
\frac{\eta(x)}{\eta(x-i\lambda)}e^{-i\lambda\partial_x}=
\lambda^{-2}e^{-i\lambda\partial_x}.
\label{142}
\end{gather}
Finally, we have
\begin{gather}
h_A^{P}=V_A{\stackrel{\circ\quad}{h_A^{P}}}V_A^{\,-1}=
\half\left(\frac{1}{\lambda^2}+i\lambda x+x^2\right)
e^{i\lambda\partial_x}
+\frac{1}{2\lambda^2}e^{-i\lambda\partial_x}=\nonumber\\
\qquad\qquad\qquad=\frac{1}{\lambda^2}
\text{ch}(i\lambda\partial_x)+
\half(x+i\lambda)xe^{i\lambda\partial_x},
\label{143}
\end{gather}
that coincides with Hamiltonian of linear relativistic
oscillator from the work \cite{AS85} (see the formula (4.1)
in this work).

\subsection{Coherent states for Meixner - Pollaczek oscillator}

We shall restrict ourself to construction of Barut - Girardello
coherent states for Meixner - Pollaczek oscillator in the space
${\stackrel{\circ\,}{\cHH^P}}$ at $\varphi=\frac{\pi}{2}.$
We have
\begin{equation}
k_-^{P}\ket{z}=z\ket{z}.\label{144}
\end{equation}
The series representation of the coherent state $\ket{z}$ by the
Fock basis
$
\left\{\ket{z}=
\wh{P}^{\nu}_{n}(\xi,\frac{\pi}{2})\right\}_{n=0}^{\infty}
$
in the space ${\stackrel{\circ\,}{\cHH^P}}$ looks like
\begin{gather}
\ket{z}=\cN^{-1}(|z|^2)\Szi{n}\frac{z^n}{\left(\mu(n-1)\right)!}
\ket{n}\,,\qquad \mu(n)=\sqrt{(n+1)(n+2\nu)}\label{145}\\
\cN^{2}(|z|^2)=\Szi{n}\frac{|z|^{2n}}{\left((\mu(n-1))!\right)^2}
=(2\nu)_n\Szi{n}\frac{|z|^{2n}}{n!\Gamma(n+2\nu)}.\label{146}
\end{gather}
The radius of convergence of the series in \reff{146} is equal
to $R=\infty.$ Taking into account \reff{051}, we obtain
\begin{equation}\label{147}
\cN^{2}(|z|^2)=\frac{\Gamma(2\nu)}{|z|^{2\nu-1}}I_{2\nu-1}(2|z|).
\end{equation}
As result we have
\begin{align}
\ket{z}&=
\frac{|z|^{\nu-\half}}{\sqrt{\Gamma(2\nu)I_{2\nu-1}(2|z|)}}
\Szi{n}(-1)^n\frac{\wh{P}^{\nu}_{n}(\xi,\frac{\pi}{2})}
{\sqrt{n!(2\nu)_n}}z^n= \nonumber \\
{}&=\frac{|z|^{\nu-\half}}{\sqrt{\Gamma(2\nu)I_{2\nu-1}(2|z|)}}
\Szi{n}(-1)^n\frac{\wt{M}^{-}_{n}(i\xi-\nu,2\nu;-1)}
{\sqrt{n!(2\nu)_n}}z^n.
\label{148}
\end{align}
Then from the relation
$$
\wt{M}^{-}_{n}(i\xi-\nu,2\nu;-1)=\sqrt{\frac{(2\nu)_n}{n!}}
M_{n}(i\xi-\nu,2\nu;-1)\,,
$$
we obtain
\begin{align}
\ket{z}&=\frac{|z|^{\nu-\half}}{\sqrt{\Gamma(2\nu)
I_{2\nu-1}(2|z|)}}\,
\Szi{n}\frac{M_{n}(i\xi-\nu,2\nu;-1)}{\sqrt{n!}}\,(iz)^n=
\nonumber \\
{}&=\frac{|z|^{\nu-\half}}{\sqrt{\Gamma(2\nu)
I_{2\nu-1}(2|z|)}}\,
e^{-iz}\,\hgs{1}{1}{i\xi-\nu}{2\nu}{-2iz}.
\label{149}
\end{align}

An overlapping of two coherent states is defined by a relation
\begin{equation}\label{150}
\braket{z_1}{z_2}=
I_{2\nu-1}(2\sqrt{\overline{z_1}z_2})
\left[I_{2\nu-1}(2|z_1|)\,I_{2\nu-1}(2|z_2|)\right]^{-\half}.
\end{equation}

\addcontentsline{toc}{chapter*}{\bf Bibliography\, .
\leaders\hbox{\, .}\hfill}\!\!\!\!

\bigskip
\bigskip

{\small\sc the St.-Petersburg University of Telecommunications,}

{\small\sc Faculty of Mathematics}

{\it e-mail:}\quad   {vadim@VB6384.spb.edu}
\medskip

{\small\sc Military Engineering Technical University,}

{\small\sc Faculty of Mathematics}

{\it e-mail:}\quad evd@pdmi.ras.ru

\end{document}